\documentclass{article}
\usepackage{amsmath,amsfonts,graphicx,epstopdf,hyperref}
\usepackage{authblk}

\begin{document}

\abovedisplayskip=15pt
\abovedisplayshortskip=15pt
\belowdisplayskip=15pt
\belowdisplayshortskip=15pt

\title{Design of quantum backflow in the complex plane}
\author{Ioannis Chremmos \\ jochremm@central.ntua.gr}
\affil{School of Electrical and Computer Engineering \\ National Technical University of Athens}

\maketitle

\begin{abstract}
A way is presented to design quantum wave functions that exhibit backflow, namely negative probability current despite having a strictly positive spectrum of momentum. These wave functions are derived from rational complex functions which are analytic in the upper half-plane and have zeros in the lower half-plane through which the backflowing behavior is controlled. In analogy, backflowing periodic wave functions are derived from rational complex functions which are analytic in the interior and have appropriately placed zeros or poles in the exterior of the unit circle. The concept is combined with a Padé-type procedure to design wave functions of this type that approximate a desired profile along the interval of backflow.
\end{abstract}

\section{Introduction}
Quantum backflow is the counter-intuitive phenomenon whereby the probability current of a quantum wave function with a strictly positive spectrum of momentum temporarily assumes negative values. More specifically, there exist quantum states $\psi(x,t)$ describing the one-dimensional motion of a free particle that contain only positive wave numbers, and yet their probability current $j(x,t)$ becomes negative in some region (or regions) of space-time. If $(x_0,t_0)$ is a point in such a region, the probability of detecting the particle in $x < x_0$ increases with time during the time interval over which $j(x_0,t)<0$. This is the opposite of what would be intuitively (or classically) expected for a forward moving particle -- or in quantum mechanical terms, a particle for which a measurement of momentum yields with certainty a positive result.

The earliest reports of quantum backflow were in the context of the time-of-arrival problem in quantum mechanics \cite{ALLCOCK_1969_III,  Kijowski1974}. This is because, under certain conditions, the probability current $j(x,t)$ of a particle wave function can be interpreted as the probability distribution of the time at which the particle arrives at the point $x$ \cite{MUGA1995}. The first analytical study of the phenomenon showed that the total amount of the backflowing probability, namely the integral of the negative probability current over time, is upper-bounded by approximately 0.04 \cite{Bracken_1994}. Remarkably, this bound is independent of the particle mass, Planck's constant and the duration of backflow, while its analytical determination remains an open mathematical problem.

From the definition of the probability current, $j = (\hbar / \mu) \: \textrm{Im} \left( \psi^* d\psi/dx \right)$, $\mu$ being the particle mass, it follows that backflow is associated with a negative local phase gradient of the wave function \cite{Berry_2010_Retro}. Hence, although its spectrum contains only positive wave numbers, the wave function oscillates in the backflow region with negative local wave numbers. From a mathematical viewpoint, backflow demonstrates that a bandlimited function can oscillate locally with frequencies that are absent from its Fourier spectrum, a property that has also drawn attention in the context of superoscillations \cite{BerrySuperOscillations} and suboscillations \cite{Chremmos_Suboscillations}.

\section{Continuous-spectrum wave functions}
\label{sec:Continuous-spectrum}
Consider a complex function defined by the ratio of two polynomials of degrees $m$ and $n$
\begin{equation}
f(z) = \frac{A_m(z)}{B_n(z)} = \frac{\displaystyle{\prod _{l=1}^{r_a}} (z-a_l)^{m_l}}{\displaystyle {\prod _{l=1}^{r_b}} (z-b_l)^{n_l}}
\label{Eq_Rational_Function}
\end{equation}
where $\{a_l\}$, $l=1,2,...,r_a,$ and $\{b_l\}$, $l=1,2,...,r_b,$ are, respectively, the distinct roots of $A_m(z)$ and $B_n(z)$ with respective multiplicities $\{m_l\}$ and $\{n_l\}$, and obviously $\sum m_l=m$ and $\sum n_l=n$. We assume that the two polynomials do not share common zeros so that all $b_l$ are poles of $f(z)$, and also $\textrm{Im}(b_l) < 0$ for all $l$. Consider also a function $\psi(x)$ of the real variable $x$, which is obtained from $f(z)$ by restricting the argument $z=x+iy$ on the real axis as
\begin{equation}
\psi(x) = N \cdot f (x+i0)
\label{Eq_Wave_Function}
\end{equation}
where $N>0$ is a normalization constant. Assuming $m < n$, we have $|\psi(x)|^2 \sim |x|^{2(m-n)}$ as $|x| \to \pm \infty$, hence $\psi(x)$ is square integrable and $N$ can be chosen to satisfy the normalization condition $\int _{-\infty} ^{+\infty} |\psi(x)|^2 dx = 1$. Then $\psi(x)$ can be regarded as a wave function representing a quantum state of a particle in position space. The quantum state can be equivalently represented in momentum space by the wave function
\begin{equation}
\tilde \psi(p) = \frac{1}{\sqrt{2\pi \hbar}} \int _{-\infty} ^{+\infty} \psi(x) e^{-ipx/\hbar} dx
\label{Eq_Momentum_Space}
\end{equation}
which is essentially the Fourier transform of $\psi(x)$. Using a standard technique based on Cauchy's theorem, the Fourier integral can be computed by integrating the complex function $f(z)\exp(-ipz/\hbar)$ along a closed path that involves the real axis and a semicircle of infinite radius in the upper or lower complex half-plane, respectively for $p<0$ or $p>0$. In either case, since $f(z) \to 0$ as $z \to \infty$, the integration over the semicircle vanishes by Jordan's lemma, leaving $\tilde \psi(p)$ equal to the sum of the residues at the poles of the integrand in the respective half-plane. By definition, the poles of $f(z)$ lie in the lower half-plane, hence $\tilde \psi(p)=0$ for $p < 0$, i.e. the quantum state represented by $\psi(x)$ has a non-negative spectrum of momentum. On the other hand, the positive part $(p>0)$ of the spectrum is derived as
\begin{equation}
\begin{aligned}
\tilde \psi(p) &= \sum _{l=1}^{r_b} P_l(p) e^{-ipb_l/\hbar}
\label{Eq_Positive_Momentum_a}
\end{aligned}
\end{equation}
where
\begin{equation}
\begin{aligned}
P_l(p) = \sum _{k=0}^{n_l-1} c_{lk} \left( -i p / \hbar \right)^k
\label{Eq_Positive_Momentum_b}
\end{aligned}
\end{equation}
and
\begin{equation}
\begin{aligned}
c_{lk} = -i K \sqrt{\frac{2 \pi}{\hbar}} \frac{f_l^{(n_l-k-1)}(b_l)}{k!(n_l-k-1)!}, \quad \quad f_l(z)  = (z-b_l)^{n_l} f(z).
\label{Eq_Positive_Momentum_c}
\end{aligned}
\end{equation}
The above equations show that each pole $b_l$ contributes to the positive spectrum a polynomial of order $n_l-1$ multiplied by the oscillating and exponentially decaying factor $\exp( -i p b_l/ \hbar)$. Notice also from \eqref{Eq_Momentum_Space} and the asymptotic behavior of $\psi(x)$ as $x \to \pm \infty$, that $\tilde \psi(0)$ exists only when $n - m > 1$, in which case $\tilde \psi (p)$ is $(n-m-2)$ times continuously differentiable at $p=0$ \cite{TrefethenSpectralMethods}. For $n - m = 1$, $\tilde \psi (p)$ is discontinuous at $p=0$ and $\tilde \psi(0)$ exists only as a principal value integral (in the sense $\lim_{L \to \infty} \int_{-L}^L$) and equals $( \tilde \psi (0-) + \tilde \psi (0+) ) / 2 = \tilde \psi (0+) / 2$. 

Now consider the probability current $j(x)$ of the quantum state $\psi(x)$ which determines the flow of probability at the instant of time when a particle is found at this state. This is expressed as
\begin{equation}
\begin{aligned}
j(x) = \frac{\hbar}{\mu} \: \textrm{Im} \left( \psi^*(x) \frac{d\psi(x)}{dx} \right) = \frac{\hbar}{\mu} \: |\psi(x)|^2  \: k(x)
\label{Eq_Probability_Current}
\end{aligned}
\end{equation}
where $^*$ denotes the complex conjugate and $k(x)$ is the local wave number
\begin{equation}
k(x) = \textrm{Im} \left( \frac{d\psi(x) / dx}{\psi(x)} \right) = \frac{d}{dx} \: \textrm{arg} \left( \psi(x) \right)
\label{Eq_Local_Wavenumber_Definition}
\end{equation}
From \eqref{Eq_Probability_Current} it is obvious that backflow, namely a negative probability current, is due to a negative local wave number. Using the definitions \eqref{Eq_Rational_Function} and \eqref{Eq_Wave_Function}
\begin{equation}
k(x) = \sum_{l=1}^{r_a} m_l\frac{ \textrm{Im}(a_l)}{|x-a_l|^2} - \sum_{l=1}^{r_b} n_l \frac{ \textrm{Im}(b_l)}{|x-b_l|^2}
\label{Eq_Local_Wavenumber_Lorentzians}
\end{equation}
This expression shows that each zero $(a_l)$ or pole $(b_l)$ of $f(z)$ contributes to $k(x)$ a Lorentzian function that is centered at $x=\textrm{Re}(a_l)$ or $\textrm{Re}(b_l)$, has a FWHM equal to $2 | \textrm{Im}(a_l)|$ or $2 |\textrm{Im}(b_l)|$ and is scaled by the respective multiplicity. A minus sign distinguishes the contribution of the poles. By assumption, $\textrm{Im} (b_l) < 0$ for all $l$, hence the poles contribute positive Lorentzians to the local wave number, thus increasing the tendency of the probability density $|\psi(x)|^2$ to move toward the positive-$x$ direction, if evolved in time. This is expected since, according to the previous analysis, poles in the lower half-plane are associated with the values of the momentum wave function $\tilde \psi(p)$ for $p>0$. On the other hand, the zeros of $f(z)$ can be placed anywhere in the complex plane without affecting the zero value of $\tilde \psi(p)$ in $p<0$ and, according to \eqref{Eq_Local_Wavenumber_Lorentzians}, zeros in the lower half-plane $(\textrm{Im} (a_l) < 0)$ contribute negative Lorentzians to $k(x)$ and thus can be used to obtain backflow.
 
\begin{figure}[t]
\includegraphics[width=1.0\textwidth]{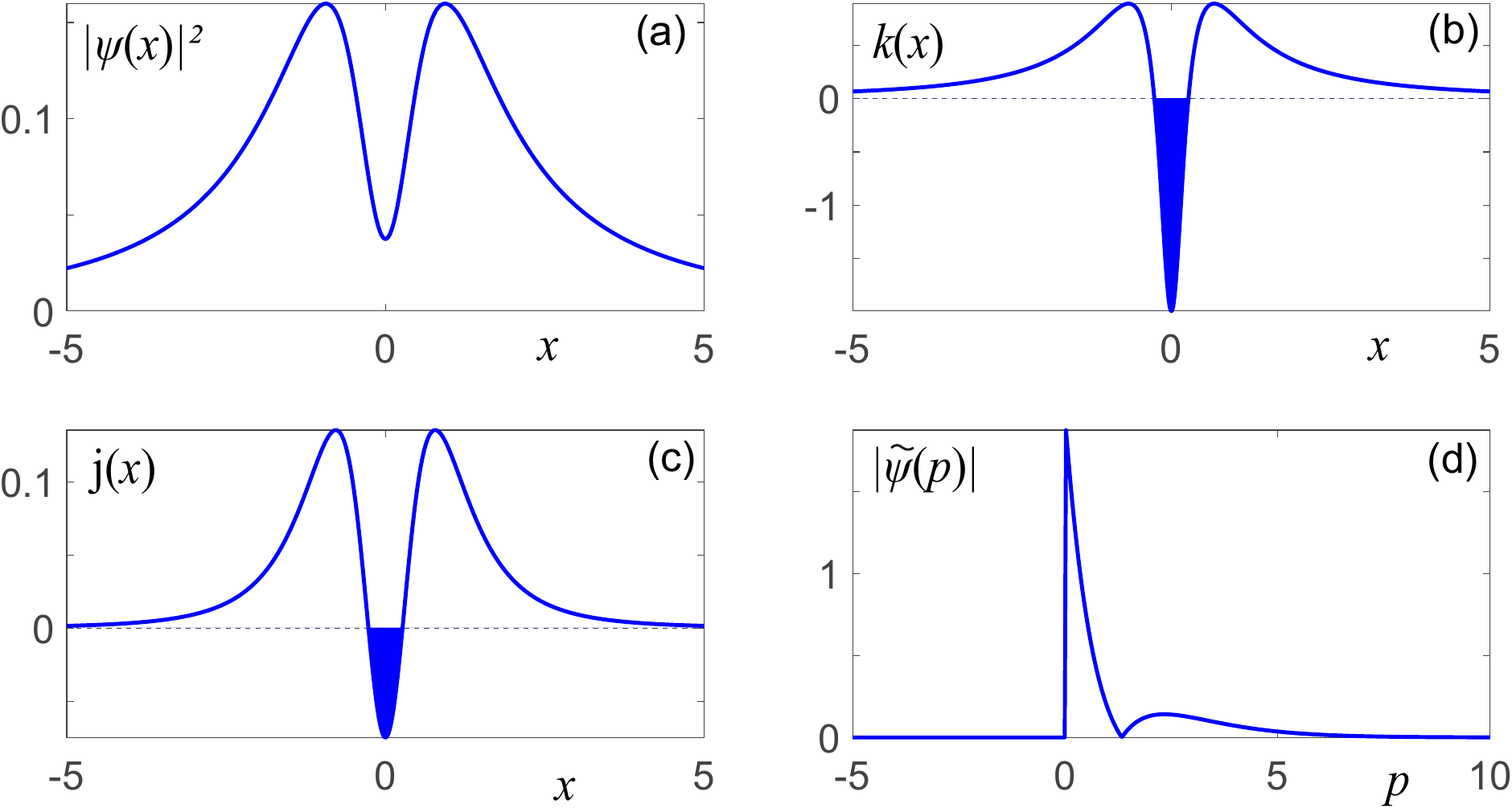}
\caption{(a) Probability density, (b) local wave number, (c) probability current and (d) spectrum of momentum of the wavefunction \eqref{Eq_Example_1_Wavefunction} for $a = -i/4$. The shaded areas in (b) and (c) indicate the interval of backflow.}
\label{fig:Example_1}
\end{figure}

To illustrate the above, we consider the simplest example of the wave functions defined through \eqref{Eq_Rational_Function} and \eqref{Eq_Wave_Function} 
\begin{equation}
\psi(x) = N \frac{x-a}{(x+i)^2}
\label{Eq_Example_1_Wavefunction}
\end{equation}
where $N = \left[ \pi(|a|^2+1)/2 \right]^{-1/2}$ and complex $a=a_1 + ia_2$ is a parameter. By direct application of the formulas \eqref{Eq_Positive_Momentum_a}-\eqref{Eq_Positive_Momentum_c}, the corresponding momentum wave function is 
\begin{equation}
\tilde \psi(p) = -i N \sqrt{2\pi} \left[ 1  + (ia-1)p \right] e^{-p}
\label{Eq_Example_1_Momentum}
\end{equation}
for $p>0$ and zero for $p<0$. Without any effect on the results, $\tilde \psi(0)$ can be assigned the principal value of the Fourier integral, which is $-iK(\pi/2)^{1/2}$. Note that, here, $x$ and $a$ are in units of an arbitrary length scale $x_0$ and $p$ is in units of $p_0=\hbar/x_0$. From \eqref{Eq_Local_Wavenumber_Lorentzians} the local wave number of this wave function is (in units of $x_0^{-1})$
\begin{equation}
k(x) = \frac{a_2}{(x-a_1)^2 + a_2^2} + \frac{2}{x^2+1}
\label{Eq_Example_1_Wavenumber}
\end{equation}
Obviously, $a$ must lie in the lower half-plane $(a_2<0)$ so that $k(x)$ obtains negative values. After some straightforward algebra, it is specifically found that $a$ must satisfy both $a_2<0$ and $|a + 5i/4| > 3/4$, namely also lie in the exterior of the circle with center $(0,-5/4)$ and radius 3/4. In this region, $k(x)=0$ has two solutions $x_1,x_2$, and $k(x)<0$ for $x_1<x<x_2$ as long as $a_2>-2$, namely we obtain backflow over a finite interval. On the other hand, if $a_2<-2$, $k(x)<0$ in the semi-infinite intervals $x<x_1$ and $x>x_2$, namely along the tails of the wave function. For $a_2=-2$, $k(x)=0$ has only one solution $x_1$, hence backflow is obtained either in $x>x_1$ (if $a_1>0$) or in $x<x_1$ (if $a_1<0$). Figure \ref{fig:Example_1} illustrates an example of the wave function \eqref{Eq_Example_1_Wavefunction}. The probability current is given in units of the frequency $\omega_0 = p_0^2/(\mu \hbar)$.

It must be noted that, although the local wave number of a backflowing wave function can assume any negative value, this is not true for the probability current which is limited by the probability density of the wave function in the interval of backflow. Indeed, the local wave number of the wave function \eqref{Eq_Example_1_Wavefunction} at $x=0$
\begin{equation}
k(0) = \frac{a_2}{|a|^2} + 2
\label{Eq_Example_1_Wavenumber_at_0}
\end{equation}
assumes unbounded negative values if $a_1=0$ and $a_2 \to 0^{-}$, while its corresponding probability current
\begin{equation}
j(0) = |\psi(0)|^2 k(0) = \frac{2}{\pi} \frac{a_2 + 2|a|^2}{1 + |a|^2},
\label{Eq_Example_1_Probability_Current_at_0}
\end{equation}
is limited from below by $(2-\sqrt{5})/\pi$, a value assumed for $a = i(2-\sqrt{5})$.

\section{Discrete-spectrum wave functions}

The concept presented in the previous section can also be applied to periodic wave functions. Consider the $L-$periodic  function obtained from $f(z)$ of \eqref{Eq_Rational_Function} by restricting $z$ on the unit circle
\begin{equation}
\psi(x) = N \cdot f (e^{i 2 \pi x/L})
\label{Eq_Wave_Function_Periodic}
\end{equation}
where $N>0$ is chosen so that the normalization condition $\int _{-L/2} ^{L/2} |\psi(x)|^2 dx = 1$ is satisfied. Such a wave function can be regarded as representing the quantum state of a particle in a one-dimensional ring with circumference $L$, e.g. a photon in a ring resonator. This wave function is represented in momentum space by the series
\begin{equation}
\psi(x) = \sum _{k} c_k \frac{e^{i p_k x /\hbar}}{\sqrt{L}} 
\label{Eq_Fourier_Series}
\end{equation}
which is essentially a Fourier series of momentum eigenstates with momentum $p_k = 2 \pi k \hbar / L$, $k$ being an integer, and coefficients
\begin{equation}
c_k = \int _{-L/2} ^{L/2} \psi_L(x) \frac{e^{-i p_k x/\hbar}}{\sqrt{L}}   dx 
\label{Eq_Fourier_Series_Coefficients_1}
\end{equation}
Replacing $\psi_L(x)$ by $f(z)$ and changing the integration variable to $z = e^{i 2 \pi x/L}$, the coefficients of \eqref{Eq_Fourier_Series_Coefficients_1} are written equivalently
\begin{equation}
c_k = \frac{\sqrt{L}}{2 \pi i} \oint _C f(z) z^{-k-1} dz 
\label{Eq_Fourier_Series_Coefficients_2}
\end{equation}
where the integral is along the unit circle $C:|z|=1$. Now assume that all poles of $f(z)$ lie in the exterior of the unit circle ($|b_l| > 1$ for all $l$) and also that $f(z)=0$. Then $f(z)z^{-1}$ is analytic in $|z| \leq 1$, and from \eqref{Eq_Fourier_Series_Coefficients_2} and Cauchy's theorem, it follows that $c_k = 0$ for $k \leq 0$, namely $\psi_L(x)$ has a positive spectrum. On the other hand, using the definitions \eqref{Eq_Rational_Function}, \eqref{Eq_Local_Wavenumber_Definition} and \eqref{Eq_Wave_Function_Periodic} the local wave number of this wave function is
\begin{equation}
k(x) = \frac{2 \pi}{L} \left( \sum_{l=1}^{r_a} m_l \frac{ 1 - |a_l|\cos(\frac{2\pi x}{L} - \textrm{arg}(a_l))}{|e^{i 2 \pi x/L}-a_l|^2} - \sum_{l=1}^{r_b} n_l \frac{ 1 - |b_l|\cos(\frac{2\pi x}{L} - \textrm{arg}(b_l))}{|e^{i 2 \pi x/L}-b_l|^2} \right)
\label{Eq_Local_Wavenumber_Periodic}
\end{equation}
The above shows that both zeros and poles outside the unit circle contribute negative values to the function $k(x)$ and this occurs over the part of the period $L$ where the corresponding numerator is negative or positive, respectively. On the other hand, zeros inside the unit circle contribute only positive wave numbers since, for $|a_l| < 1$, the corresponding numerator is positive for all $x$.

\begin{figure}[t]
\includegraphics[width=1.0\textwidth]{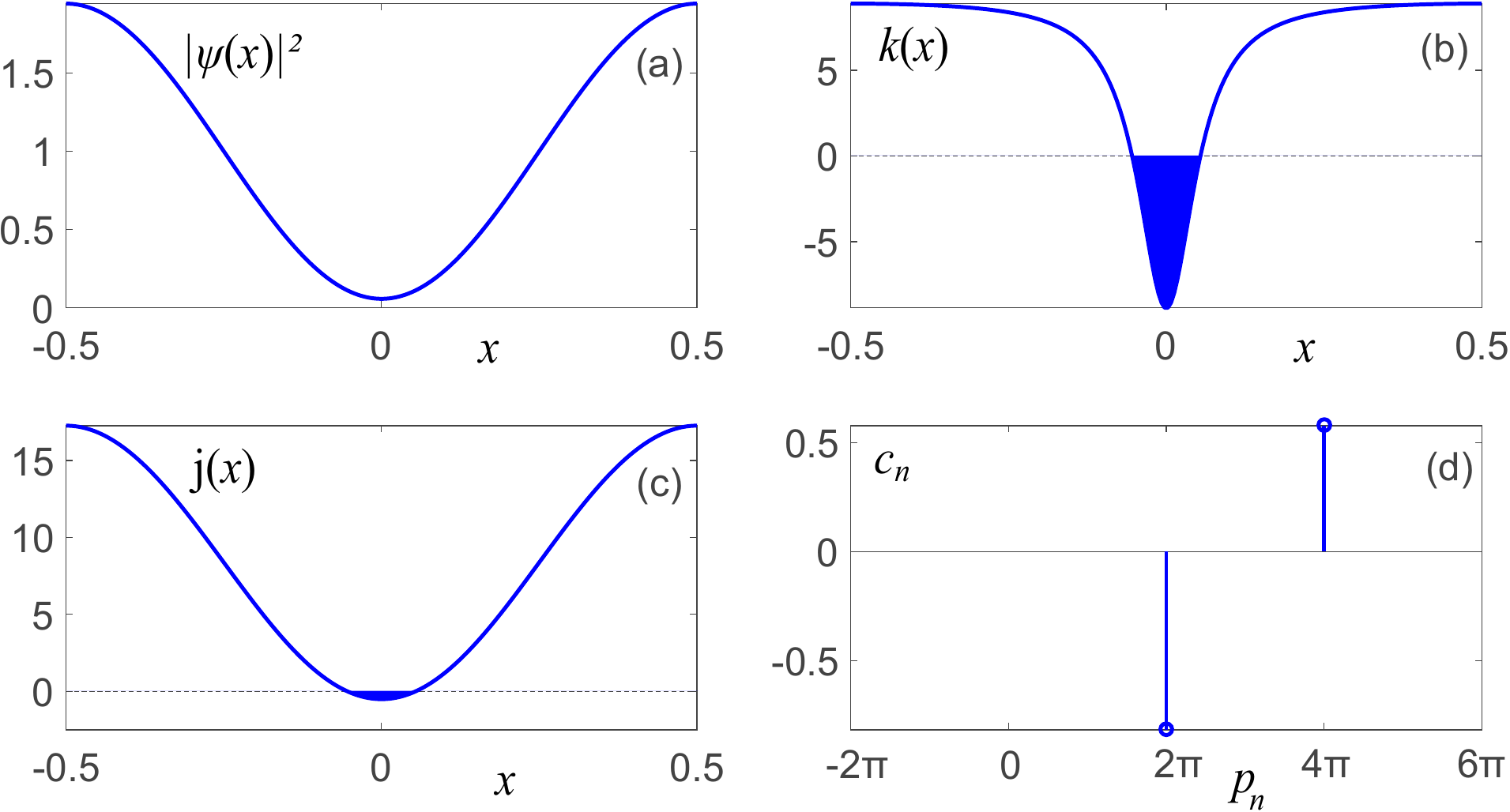}
\caption{((a) Probability density, (b) local wave number, (c) probability current and (d) spectrum of momentum in a period of the wavefunction \eqref{Eq_Example_2_Wavefunction} for $a = \sqrt{2}$. The shaded areas in (b) and (c) indicate the interval of backflow.}
\label{fig:Example_2}
\end{figure}

As an example, consider the simplest case of the defined wave functions
\begin{equation}
\psi(x) = N e^{i 2\pi x} ( e^{i 2\pi x}-a )
\label{Eq_Example_2_Wavefunction}
\end{equation}
where, without loss of generality, $a$ is assumed to be real, $a>1$, and $N = (1+a^2)^{-1/2}$. Obviously, this wave function contains only two momentum eigenstates with positive momenta $2 \pi$ and $4 \pi$, while its local wave number
\begin{equation}
k(x) = 2 \pi \left( 1 + \frac{ 1 - a\cos(2 \pi x)}{1 + a^2 - 2a \cos(2 \pi x)} \right)
\label{Eq_Example_2}
\end{equation}
becomes negative for $\cos(2 \pi x) > (a^2+2)/3a$ provided that $a < 2$. Here, $x$ is in units of an arbitrary period length $L$, wave number and momentum are respectively in units of $L^{-1}$ and $\hbar/L$, while $a$ is dimensionless. Figure \ref{fig:Example_2} illustrates an example of the wave function \eqref{Eq_Example_2_Wavefunction}.

\begin{figure}[t]
\includegraphics[width=1.0\textwidth]{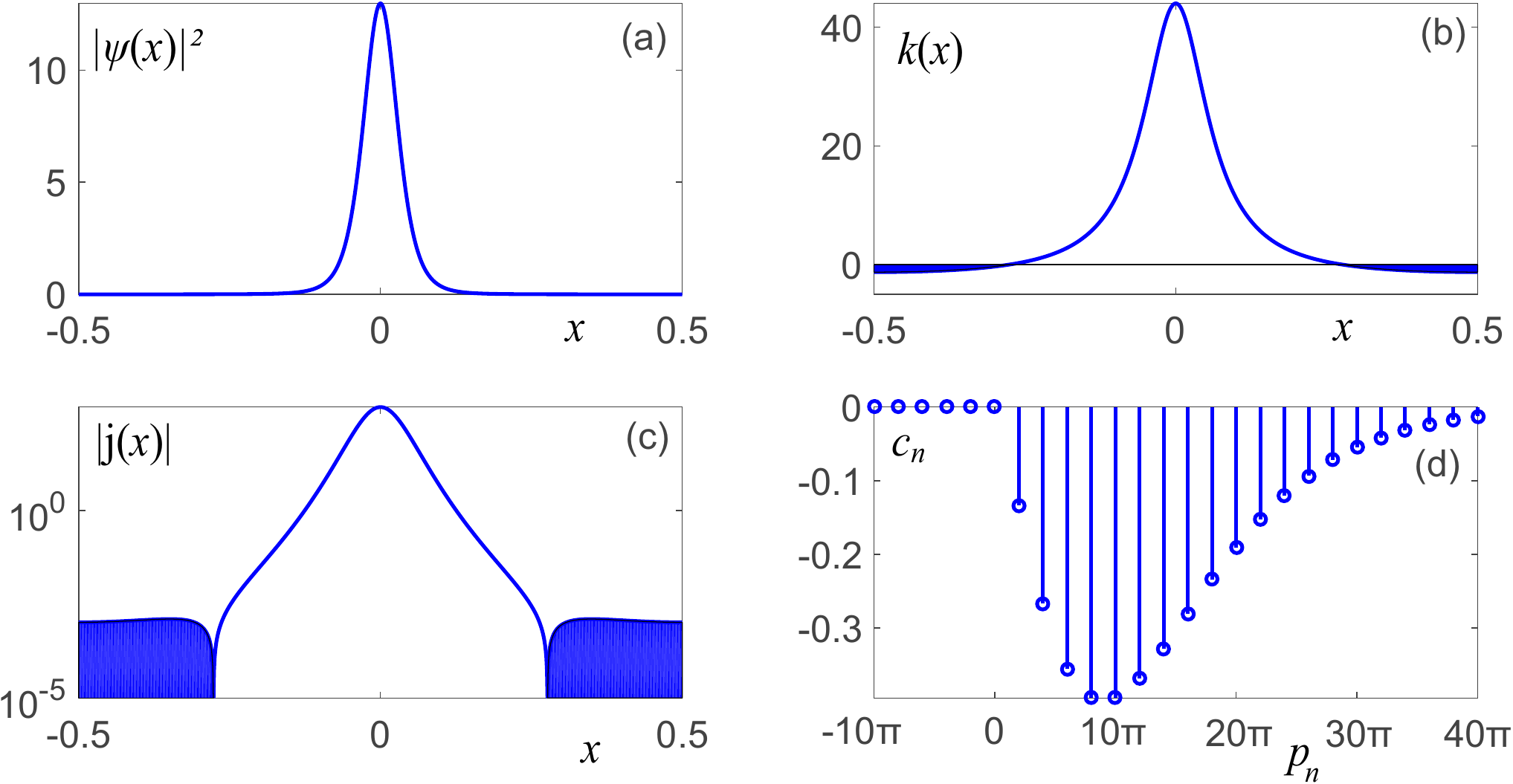}
\caption{(a) Probability density, (b) local wave number, (c) magnitude of the probability current (in logarithmic scale) and (d) spectrum of momentum in a period of the wavefunction \eqref{Eq_Example_3} for $a = 1.5$, $n=3$. The shaded areas in (b) and (c) indicate the interval of backflow.}
\label{fig:Example_3}
\end{figure}

Another simple example is when, in addition to the zero at $z=0$, $f(z)$ has a multiple pole at $z=a > 1$
\begin{equation}
\psi(x) = N \frac{e^{i 2\pi x}}{ \left( e^{i 2\pi x}-a \right)^{n} }
\label{Eq_Example_3}
\end{equation}
where the normalization constant is given by a certain integral.\footnote{$ N = (a-1)^n \sqrt{\frac{\pi}{2 c I}}$, where $I = \int_{0}^{\infty} \frac{(1 + c^2 t^2)^{n-1}}{(1+t^2)^n}dt$ and $c = \frac{a-1}{a+1}$} This function has an infinite positive spectrum of momenta $2\pi, 4\pi, ...,$ which is easily determined by a Taylor expansion with respect to $e^{i 2 \pi x}/a$
\begin{equation}
\psi(x) = \frac{-N}{(-a)^{n-1}} \sum _{k=1}^{\infty} \binom{n+k-2}{k-1} \frac {e^{i 2 \pi k x}}{a^k}
\label{Eq_Example_3_Fourier_Series}
\end{equation}
After \eqref{Eq_Local_Wavenumber_Periodic}, the local wave number of this wave function is 
\begin{equation}
k(x) = 2 \pi \left( 1 - n \frac{ 1 - a\cos(2 \pi x)}{1 + a^2 - 2a \cos(2 \pi x)} \right)
\label{Eq_Example_3_Wavenumber}
\end{equation}
and it is easily shown that it becomes negative only when $n > a+1$. Figure \ref{fig:Example_3} illustrates an example of the wave function \eqref{Eq_Example_3}.

\section{Backflow with a desired profile}

In this section we show how the concept of Section \ref{sec:Continuous-spectrum} and a procedure that derives from the theory of Padé approximants \cite{PadeApprox_Baker} can be combined to design wave functions that approximate a desired profile along the interval of backflow. Assume an interval $(-x_0,x_0)$, $x_0>0$ and a (complex) profile $N p(x)$ to be approximated, $N$ being a normalization constant. In the standard Padé procedure one would determine the polynomials $A_m(x)$ and $B_n(x)$ so that the Taylor series of their ratio matches that of $p(x)$ up to order $m+n$. However this procedure does not guarantee that the poles of $B_m(z)$ lie in the lower half-plane, which is a prerequisite for the wave function to have a strictly positive spectrum of momentum. For this reason we first choose freely the poles of the polynomial $B_n(z)$ and subsequently use the Padé procedure to determine $A_m(x)$ so that the Taylor series of $A_m(x)/B_n(x)$ matches that of $p(x)$ up to order $m$. If
\begin{equation}
A_m(x) = \sum_{k=0}^{m} \alpha_k x^k, \qquad B_n(x) = \sum_{k=0}^{n} \beta_k x^k
\label{Eq_Polynomials_Series}
\end{equation}
are the power forms of the corresponding polynomials and $\sum_{k=0}^{\infty} p_k x^k$ is the Taylor series of $p(x)$, the described procedure leads to the coefficients $\alpha_k$ being expressed in terms of the coefficients $\beta_k$ and $p_k$ as
\begin{equation}
\alpha_k = \sum_{l=0}^{k} \beta_{l} p_{k-l}
\label{Eq_Pade_Coefficients}
\end{equation}
for $k=0,1,...,m$. After normalization with the $N$ factor (which is generally determined numerically), we finally obtain the wave function
\begin{equation}
\psi(x) = N \frac{A_m(x)}{B_n(x)} = N \sum_{k=0}^m p_k x^k + O(x^{m+1})
\label{Eq_Pade_Rational_Function}
\end{equation}
whose Taylor series has the same first $m+1$ terms with the profile function $N p (x)$. Obviously, the free choice of $B_n(z)$ implies that there is no unique solution to the stated approximation problem. However, it can be shown that the magnitude of the error term $O(x^{m+1})$ in \eqref{Eq_Pade_Rational_Function} decreases with increasing relative magnitude of the poles of $B_n(z)$ and the length $(2x_0)$ of the interval of backflow, which implies that a good approximation limits the choice of the poles to values $|b_l| \gg x_0$.

\begin{figure}[t]
\includegraphics[width=1.0\textwidth]{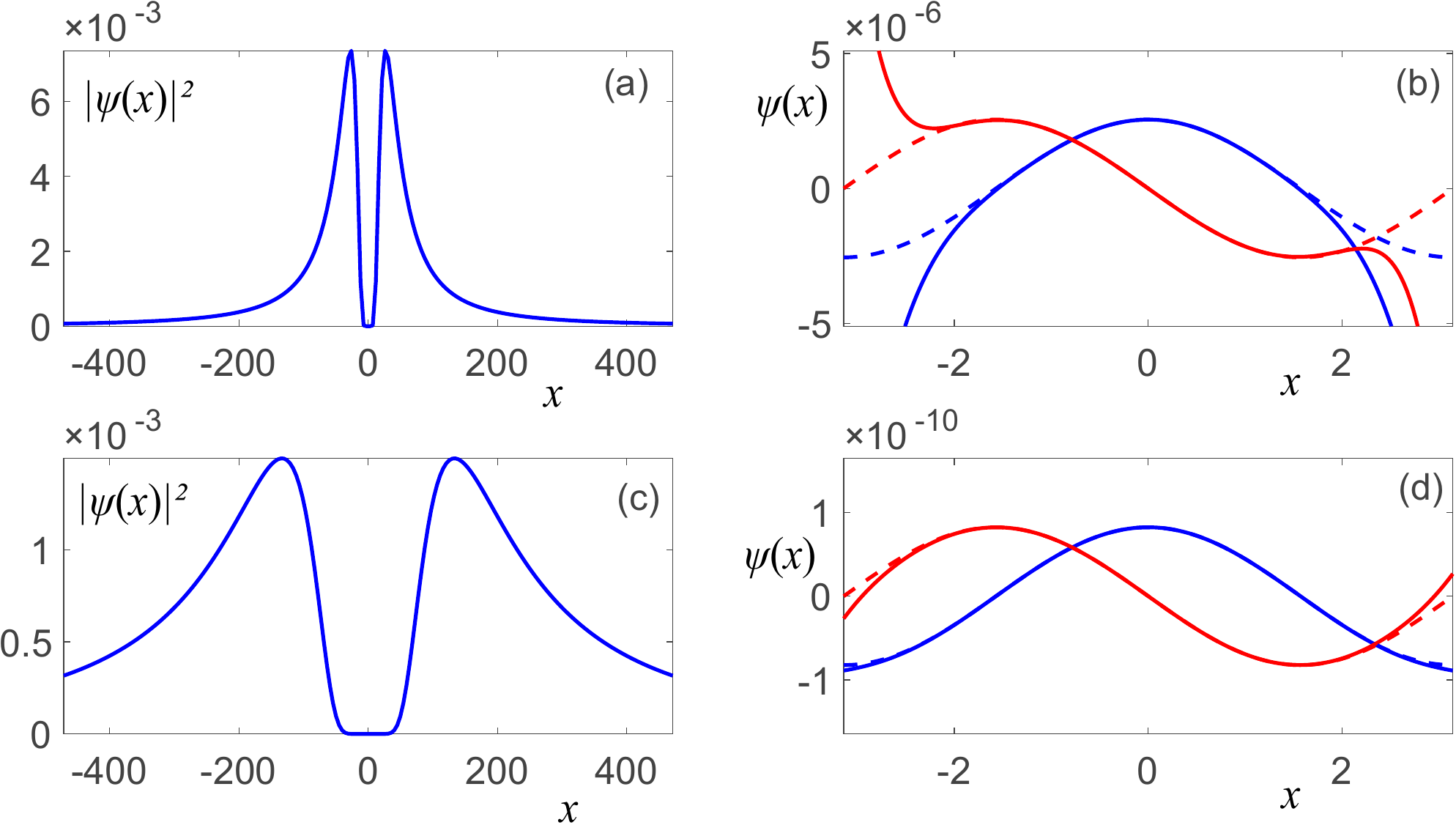}
\caption{(a) Probability density and (b) real (blue line) and imaginary (red line) part of the wave function \eqref{Eq_Example_4_Wavefunction} for $x_0 = \pi$, $m=8$ and $b = 3\pi$. The dashed lines correspond to $Ne^{-ix}$. (c-d) Same as (a-b) with $b = 15\pi$.}
\label{fig:Example_4}
\end{figure}

As an example consider the wave function
\begin{equation}
\psi(x) = N  \frac{\sum_{k=0}^{m} \alpha_k x^k}{(x + ib)^{m+1}}
\label{Eq_Example_4_Wavefunction}
\end{equation}
and the backflowing profile $p(x) = e^{-ix}$ to be approximated in an interval $(-x_0,x_0)$, $x_0>0$. The denominator polynomial has been chosen arbitrarily to have a single pole of multiplicity $m+1$ at $z=-ib$, $b>0$. Substistuting in \eqref{Eq_Pade_Coefficients} the expressions of coefficients $\beta_k$ and $p_k$ we obtain
\begin{equation}
\alpha_k = \sum_{l=0}^k \binom{m+1}{l} \frac{(ib)^{m+1-l} (-i)^{k-l}}{(k-l)!}
\label{Eq_Pade_Example_coefficients}
\end{equation}
for $k=0,1,...,m$. Figure \ref{fig:Example_4} shows two examples of the wave function \eqref{Eq_Example_4_Wavefunction} for $x_0 = \pi$, $m = 8$ and $b = 3\pi$ or $b = 15\pi$. Notice the dramatic increase of the amplitude of the wave function outside the interval of backflow. The ratio of the maximum amplitude over the amplitude in the interval of backflow can be shown to scale approximately as $b^m / m!$, namely it increases (for large $b$) with the length of the interval (a larger $x_0$ implies larger $m$ and $b$) and with the accuracy of the approximation (a larger $b$ is required for smaller error). Indeed, for $b=15x_0$ the wave function approximates the $e^{-ix}$ profile in $(-\pi, \pi)$ much better compared to $b=3x_0$ however at the cost of a $~10^{5}$ times lower amplitude.

The low probability density of a backflowing wave function in the interval of backflow is a general attribute of all functions that oscillate over finite intervals with local wave numbers that are absent from their spectrum, such as superoscillatory and suboscillatory functions, and remains the main challenge toward their practical applications \cite{Berry_2019_Roadmap}.

\section{Conclusions}

A systematic method for designing backflowing wave functions was presented for the first time to our knowledge. Such functions were obtained from rational complex functions which are analytic in the upper half-plane and have zeros in the lower half-plane to impart locally a negative phase gradient. Backflowing periodic wave functions were similarly obtained from rational complex functions which are analytic in the interior and have critically placed zeros or poles in the exterior of the unit circle. A Padé-type procedure was demonstrated to design wave functions of this type that approximate a desired profile along the interval of backflow with arbitrary degree of accuracy.

\bibliography{mypapers}
\bibliographystyle{ieeetr}

\end{document}